\documentclass{seg}
\usepackage{bm}
\usepackage[authoryear,round]{natbib}
\usepackage{booktabs}
\hypersetup{
    colorlinks=true,
    citecolor=blue,
    linkcolor=blue,
    urlcolor=blue
}
\title{A general lightweight global modeling framework for three-dimensional seismic exploration}

\author{%
  Changxin Wei$^{1}$\and
  Jun Ma$^{1}$\and
  Xintong Dong$^{1,*}$\thanks{Corresponding author: dxt@jlu.edu.cn}
}
\date{14 September 2026}

\begin{document}

\maketitle

\begin{center}
    \footnotesize
    $^{1}$ State Key Laboratory of Deep Earth Exploration and Imaging, College of Instrumentation and Electrical Engineering, Jilin University, Changchun, China\\
\end{center}

\begin{abstract}
In seismic exploration, the propagation of seismic waves naturally gives rise to long-range dependencies in seismic data. Capturing such global correlations can significantly improve the accuracies of seismic signal processing, inversion, and interpretation. Global modeling (GM) methods have therefore emerged as an effective paradigm for seismic exploration, offering a powerful means of exploiting the intrinsic global relationships within seismic data. However, the mainstream GM approaches, particularly Transformers, incur prohibitively high computational costs for their inherent quadratic complexity, which restricts the scalability of such methods to higher-dimensional data. To achieve efficient GM for three-dimensional (3D) seismic exploration, we propose a general framework, named lightweight Mamba-based global modeling (LMGM), to establish long-range dependencies at a relatively low computational cost. Specifically, LMGM fully leverages the linear complexity of Mamba to circumvent the prohibitive computational cost of higher-dimensional data. A three-directional scanning Mamba (3DSM) block is designed to exploit spatial correlations along the three dimensions of 3D seismic data. Meanwhile, a dual-domain-aware (DDA) block is further developed to integrate time- and frequency-domain features for enhanced feature representation. In addition, a two-stage plug-and-play nonlinear normalization (TPNN) strategy is proposed to enhance the adaptability of LMGM to diverse seismic records while preserving signal characteristics. We employ 3D seismic data interpolation as a representative task to comprehensively evaluate LMGM on both synthetic and field data. The results demonstrate that LMGM achieves superior processing performance while substantially reducing computational cost compared with U-Net-based and Transformer-based methods. Furthermore, experiments on random noise removal demonstrate the applicability of LMGM beyond interpolation, supporting its potential as a general GM framework for a variety of 3D seismic data processing tasks.
\end{abstract}

\section{Introduction}
Seismic signals essentially represent the spatiotemporal responses of the subsurface medium to source excitation. The propagation of seismic waves can induce strong spatial and temporal correlations across traces and sampling points. \citep{campillo2003long, qin2026adaptive} Therefore, long-range dependencies intrinsically exist in seismic data. This property makes global modeling (GM) a natural choice for seismic data, as it enables the exploitation of their inherent global structure.

In the past decade, GM has been widely adopted in the fields of natural language processing \citep{vaswani2017attention, wolf2020transformers, gillioz2020overview, gu2023mamba, qu2026survey} and computer vision \citep{han2022survey, liu2025vision}. Compared with conventional convolutional neural networks (CNNs), which are structurally constrained by limited receptive fields \citep{wang2018non}, GM can establish substantially larger receptive fields by jointly consider all elements within the input, allowing effective extraction of long-range dependencies and better characterization of complex patterns.

These advantages have also been increasingly employed in seismic exploration. For example, in denoising tasks, GM can further enhance the continuity of seismic events \citep{chen2024efficient}; in interpolation tasks, GM provides more correlations for reconstructing consecutively missing traces \citep{cheng2024seismic}; in super-resolution tasks, GM can better characterize multiple geotectonic features \citep{sun2025removing}. A representative architecture, Transformer \citep{vaswani2017attention}, has become a mainstream GM approach in seismic exploration. It employs multi-head self-attention (MSA) mechanism to calculate the correlations between each element and all other elements in the input through attention operations. Recently, geophysical scholars have introduced several Transformer-based frameworks into different tasks of seismic data processing \citep{wang2023seismic, jiang2023seismic, park2024transformer, wang2024self, dong2025deep, geng2026robust, zhang2026free}. For instance, in seismic velocity inversion tasks, it is challenging to recover deep subsurface structures due to the weaken reflection signals, and multiples generated by some complex subsurface structures will cause significant interference in seismic records \citep{wang2023seismic, li2025towards}. These issues make GM particularly important for velocity inversion. Motivated by this, \cite{wang2023seismic} were the first to apply Transformer to seismic velocity inversion and achieved greater results than CNN-based methods. In the next year, \cite{wang2024self} further proposed seismic data denoising Transformer with a novel self-supervised pre-training strategy, demonstrating promising effectiveness and flexibility in attenuating various complex seismic noises. Different from regularly or randomly missing cases, the consecutively missing case needs to capture more correlations from distant seismic traces. However, the conventional local operator, i.e., CNNs, can hardly capture long-range dependencies from such big gaps. To overcome this limitation, \cite{cheng2024seismic} proposed a seismic interpolation Transformer, which adopts an encoder-decoder architecture and incorporates U-shaped Swin-Transformer \citep{liu2021swin} blocks in the bottleneck. Both comparative and ablation experiments demonstrate the significance of GM for such challenging interpolation task. \cite{wu2025robust} employed a physics-informed Transformer-based autoencoder network for full waveform inversion in an unsupervised manner, achieving higher accuracy with lower parameters compared to a CNN-based method and a Transformer-only method.

Nevertheless, the quadratic computational complexity of the Transformer poses a major challenge. Given an input of length $n$, the computational complexity is $O(n^2)$ since self-attention computes pairwise interactions among all elements to capture their relationships. As a result, the computational cost will increase dramatically with the data size. Therefore, most of the aforementioned methods are tailored to two-dimensional (2D) seismic data with relatively moderate sizes. When extending Transformer-based methods to three-dimensional (3D) seismic data with relatively larger sizes, the challenge becomes more severe. Only a few studies have made such attempts that employ simplified or modified Transformers, so as to mitigate or avoid the high computational cost. For instance, \cite{jiang20243} constructed a Swin-Transformer-based 3D first break picking network with a two-channel mask strategy, while achieving greater performance than 3D U-Net at the cost of nearly twice the processing time. \cite{wang2025hcvt} proposed a hybrid CNN-Transformer network for 3D seismic data interpolation by incorporating an improved vision Transformer \citep{dosovitskiy2020image} at the bottleneck of U-Net. Here, the MSA is computed only on low-resolution feature representations rather than full-resolution ones, thereby elegantly avoiding the high computational cost when applying Transformer to 3D data. Despite their low computational cost, these 3D applications have not fully exploited the potential of GM. These limitations call for a more computationally efficient GM framework in 3D seismic exploration without sacrificing the modeling capability.

Recently, Mamba \citep{gu2023mamba} has emerged as a GM architecture and attracted increasing attention \citep{liu2025vision, qu2026survey}. Mamba is based on a kind of state space model (SSM) enhanced by a selective scan mechanism and a hardware-aware algorithm, allowing the model to selectively propagate or forget information and substantially reducing memory swaps \citep{gu2023mamba}. In Mamba, the SSM selectively scans the input sequence and recursively updates its hidden state by combining the current input with the previous state. The information is progressively propagated across the sequence, thereby establishing long-range dependencies. Since the SSM processes each input element sequentially instead of computing pairwise interactions, the computational complexity of Mamba is $O(n)$ for a sequence with a length of $n$. Compared to Transformer, Mamba achieves acceleration at both the hardware and software levels, thereby allowing it to be considered as a lightweight GM architecture and a promising alternative to Transformer. Consequently, Mamba has been successfully applied to image restoration \citep{guo2024mambair}, image classification \citep{yue2024medmamba}, and image segmentation \citep{ruan2024vm}, demonstrating remarkable GM capability and high efficiency for 2D images. Benefiting from these advantages, geophysical scholars have also attempted to introduce Mamba into seismic data processing. For example, \cite{chen2025dual} proposed to incorporate Mamba with fast Fourier convolutions to form a frequency-augmented state space module, which attenuates seismic random noise in both time domain and frequency domain. Experimental results demonstrate its superior denoising performance with substantially lower computational cost compared to Transformer-based methods. Moreover, in our previous work \citep{wei2026globally}, we have also proven Mamba’s efficiency in seismic data interpolation by designing an interpolation network which exploits the collaborative advantages of Mamba and dense connections \citep{huang2017densely} in a lightweight architecture. These successful and efficient applications to 2D seismic data demonstrate the potential of Mamba for globally modeling 3D seismic data. To our best knowledge, no prior work has made such attempts. Therefore, we propose a lightweight Mamba-based GM framework (LMGM) for 3D seismic data processing in this paper. Specifically, the proposed LMGM framework is characterized by the following key designs: 1) Considering the long-range dependencies among the three dimensions of 3D seismic data, we design three-directional scanning Mamba (3DSM) block to better exploit 3D spatial correlation information; 2) Considering the rich frequency-domain characteristics of seismic data, we design a dual-domain-aware (DDA) block to further enhance the extracted global features, with two separate branches processing time- and frequency-domain information, respectively; 3) For some seismic records containing large amplitude variations to recover, we propose a two-stage plug-and-play nonlinear normalization (TPNN) strategy to compress the data range and boost the amplitude of late arrivals. We apply LMGM to 3D interpolation experiments on both synthetic and field data as a representative task to validate its effectiveness and efficiency. Additionally, in the discussion part, we further investigate the performance of LMGM on seismic data denoising tasks. The results demonstrate that LMGM has the potential to serve as a general GM framework for a variety of 3D seismic data processing tasks.

\section{Methodology}
In this section, detailed descriptions of LMGM, 3DM block and 3DSM block, Mamba module, DDA block and TPNN are provided. The shape of the input 3D data patches is set to 32×32×8. Taking interpolation as example, the input and output of LMGM are illustrated in Fig. \ref{fig1}a as incomplete and recovered 3D volumes, respectively.

\subsection{Overall network architecture of LMGM}
\begin{figure}[htbp]
    \centering
    \includegraphics[width=0.7\textwidth]{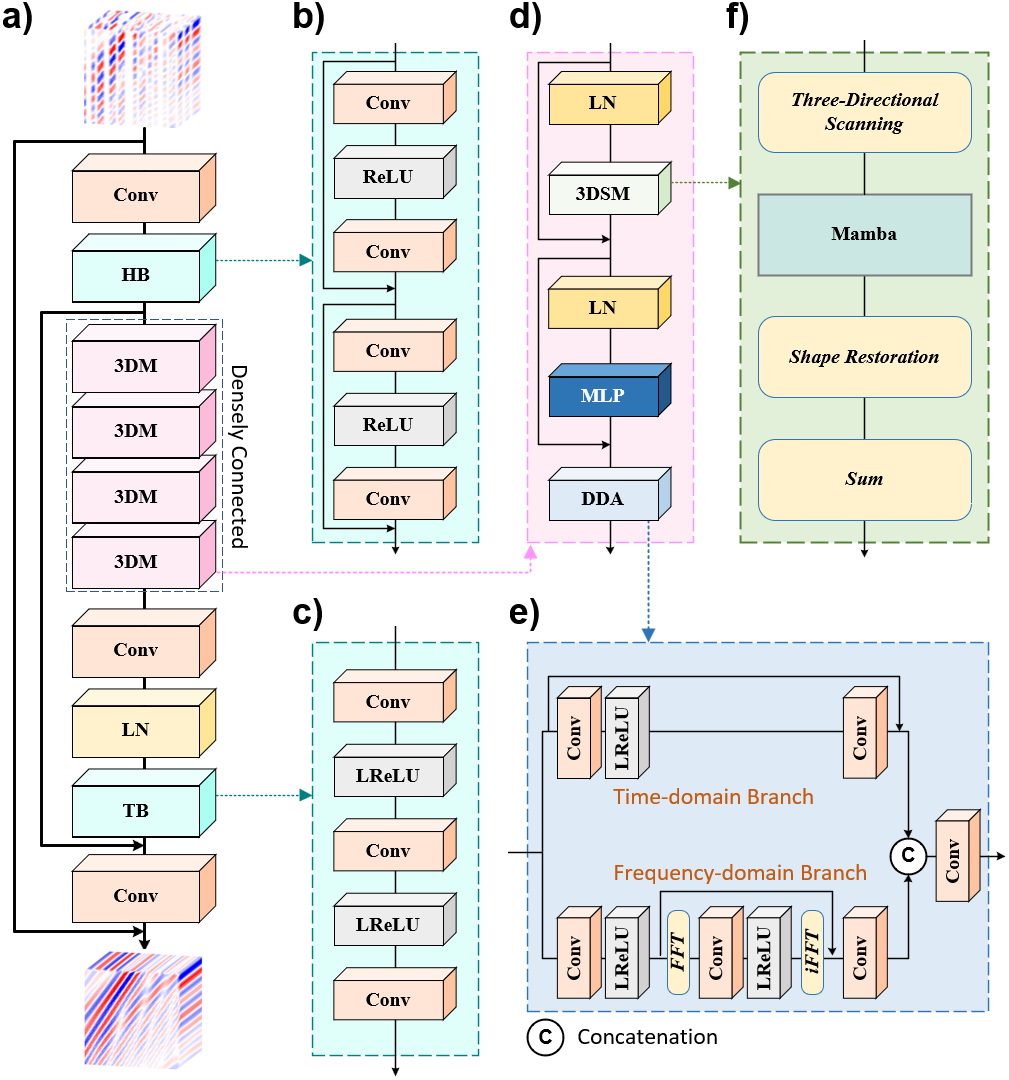}
    \caption{Network architectures of (a) LMGM, (b) HB, (c) TB, (d) 3DM block, (e) DDA block, and (f) 3DSM block, respectively.}
    \label{fig1}
\end{figure}

As shown in Fig. \ref{fig1}a, the corrupted data is firstly propagated into a 3D convolution layer with a kernel of 3×3×3 (denoted as Conv), and its channel number is increased from 1 to 96. Then, it is passed through a head block (HB) to extract initial features. As shown in Fig. \ref{fig1}b, an HB is composed of two identical parts, each comprising two Convs, a rectified linear unit (ReLU), and a residual connection. Subsequently, the output of HB is fed into the core GM module, which consists of four consecutive 3D Mamba (3DM) blocks. The four 3DM blocks are densely connected \citep{huang2017densely} to enhance the global feature reuses and interactions. Then, a Conv is used to refine the global feature, and a layer normalization (LN) layer is employed to stabilize the training process. The following tail block (TB) is composed of three Convs interleaved with leaky ReLUs (LReLU) in a feed-forward manner as shown in Fig. \ref{fig1}c, which is deployed to enable non-linear transformation while avoiding neuron inactivation. The output features of TB and HB are fused via a residual connection, thus avoiding gradient vanishing. At the end of the network, the last Conv is used to integrate features across different channels and reduce the channel number back to 1. Meanwhile, another residual connection directly transfers the input of the network to the output.

\subsection{3DM block and 3DSM block}
As illustrated in Fig. \ref{fig1}d, the 3DM block follows the Pre-LN \citep{xiong2020layer} architecture, where LN layer is placed before each feature transformation module (i.e., 3DSM or multi-layer perceptron (MLP)) within the residual branch. Meanwhile, a DDA block is placed at the end of the 3DM block to enhance the extracted global features. The workflow of the 3DM is as follows:
\begin{equation}
    \bm{X}_1=3DSM(LN(\bm{X}))+\bm{X},
\end{equation}
\begin{equation}
    \bm{Y}=DDA(MLP(LN(\bm{X}_1))+\bm{X}_1),
\end{equation}
where $\bm{X}$ and $\bm{Y}$ denote the input and output features of the 3DM block, respectively.

\begin{figure}[htbp]
    \centering
    \includegraphics[width=1\textwidth]{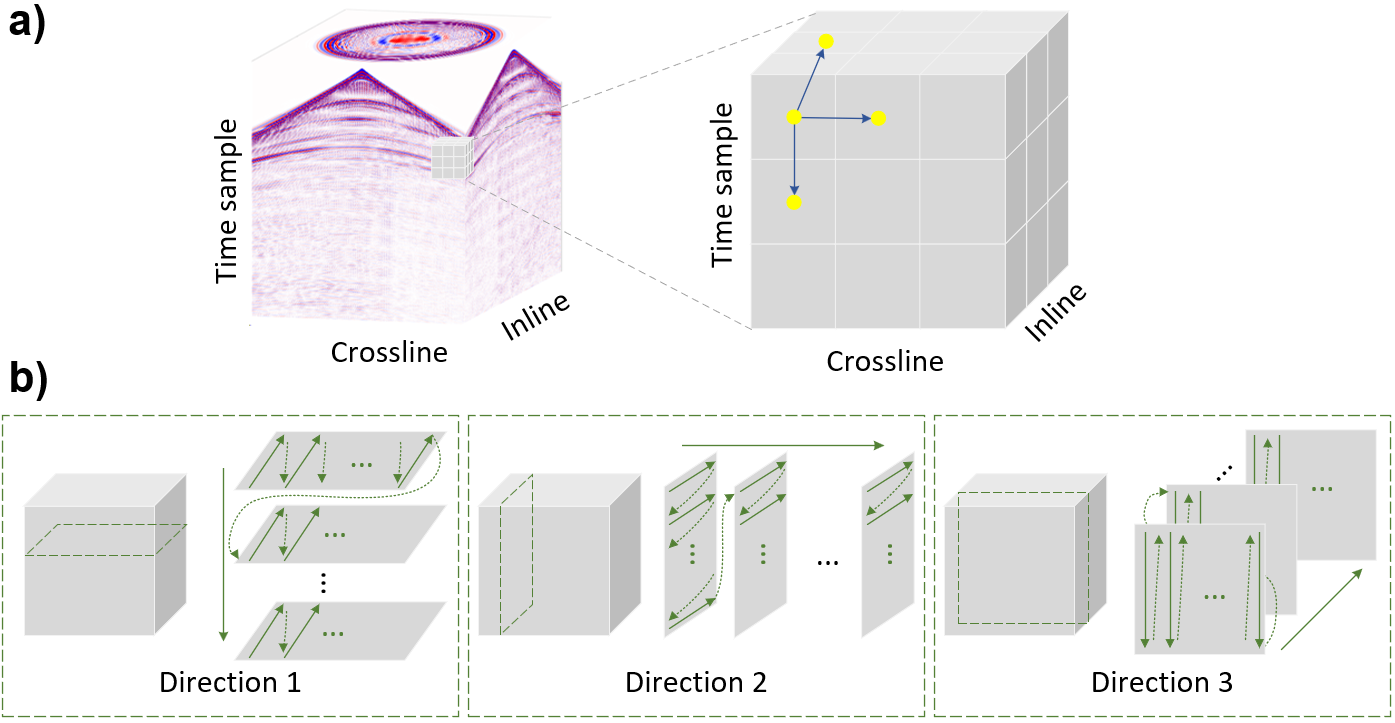}
    \caption{Illustrations of (a) the correlations among the three dimensions, and (b) the three scanning directions in 3DSM block.}
    \label{fig2}
\end{figure}

For 3D seismic data, spatial correlations exist not only between time samples and traces as 2D relationships, but also among time samples, crosslines, and inlines to form 3D relationships, as illustrated in Fig. \ref{fig2}a. To jointly leverage long-range dependencies from the three dimensions, we design the 3DSM block as in Fig. \ref{fig1}f, so as to jointly consider the correlations among sample points through different directions. In a 3DSM block, an input 3D volume is of shape ${S}\times{X}\times{I}$, where $S$ is the time samples, $X$ denotes the crosslines, and $I$ represents the inlines. This 3D feature is scanned along three directions and flattened into three sequences, i.e., $SXI$, $XSI$, and $IXS$, following the rule that the last dimension varies first. These three sequences correspond to the three directions illustrated in Fig. \ref{fig2}b, respectively.

Subsequently, the three sequences are input into a reusable Mamba module, after which their shapes are restored to the same as the original input. The three features are then element-wise summed to generate the final output of the 3DSM block. Given an input feature $\bm{X}$, the output feature $\bm{Y}$ of 3DSM is obtained as follows:
\begin{equation}
\bm{S}_1,\bm{S}_2,\bm{S}_3=3DS(\bm{X}),
\end{equation}
\begin{equation}
\bm{Y}_1^\prime=Mamba(\bm{S}_1),\bm{Y}_2^\prime=Mamba(\bm{S}_2),\bm{Y}_3^\prime=Mamba(\bm{S}_3),
\end{equation}
\begin{equation}
\bm{Y}_1=SR(\bm{Y}_1^\prime),\bm{Y}_2=SR(\bm{Y}_2^\prime),\bm{Y}_3=SR(\bm{Y}_3^\prime),
\end{equation}
\begin{equation}
\bm{Y}=\bm{Y}_1+\bm{Y}_2+\bm{Y}_3,
\end{equation}
where $\bm{S}_1$, $\bm{S}_2$, and $\bm{S}_3$ represent the sequences scanned along the three directions, $3DS$ denote the three-directional scanning, $\bm{Y}_1^\prime$, $\bm{Y}_2^\prime$, and $\bm{Y}_3^\prime$ denote the direct outputs of the Mamba, $\bm{Y}_1$, $\bm{Y}_2$, and $\bm{Y}_3$ are the corresponding outputs after shape restoration, and $SR$ is the shape restoration operation.

\subsection{Mamba module}
\begin{figure}[htbp]
    \centering
    \includegraphics[width=0.4\textwidth]{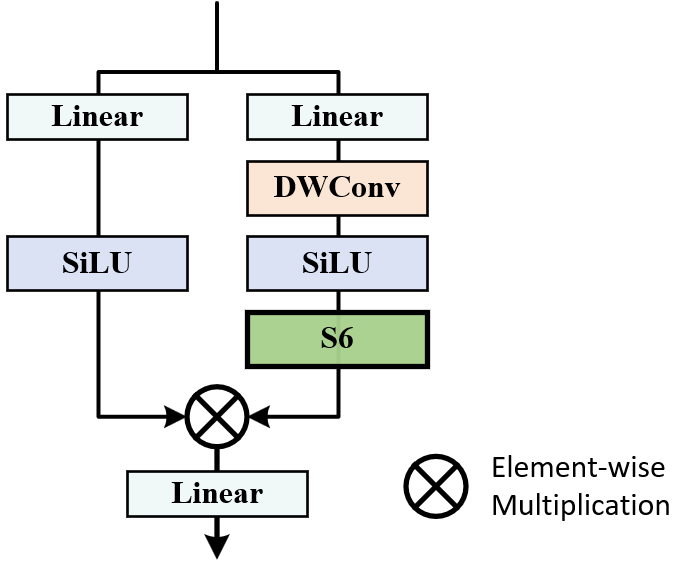}
    \caption{The Mamba architecture.}
    \label{fig3}
\end{figure}

The architecture of the Mamba module is illustrated in Fig. \ref{fig3}, where the input feature is first projected and divided into two paths. The main path is processed by a linear layer, depth-wise convolution (DWConv), and a sigmoid linear unit (SiLU) activation before being fed into the S6 algorithm \citep{gu2023mamba}, which refers to the SSM enhanced with selective mechanism and computed with a scan. Here, the long-range dependencies are captured through the iterations of the SSM. Meanwhile, the other path serves as a gating mechanism through the SiLU activation. The outputs of the two paths are fused via element-wise multiplication, followed by a linear projection to obtain the final output. The calculation workflow within the Mamba module is formulated as follows:
\begin{equation}
\bm{X}_1=S6(SiLU(DWConv(Linear(\bm{X})))),
\end{equation}
\begin{equation}
\bm{X}_2=SiLU(Linear(\bm{X})),
\end{equation}
\begin{equation}
\bm{Y}=Linear(\bm{X}_1\odot\bm{X}_2),
\end{equation}
where $\bm{X}$ and $\bm{Y}$ are the input and output features of the Mamba module, respectively, and $\odot$ denotes the element-wise multiplication.

\subsection{DDA block}
As shown in Fig. \ref{fig1}e, a DDA block consists of two parallel branches for jointly modeling time- and frequency-domain features. In the time-domain branch, a Conv, a LReLU, and another Conv with a residual connection are employed to enhance local representations of input features. In the frequency-domain branch, the features are first propagated through a Conv and a LReLU. Then, it is transformed into the frequency domain using the fast Fourier transform (FFT), followed by a Conv and a LReLU for frequency-domain feature extraction. The result is then transformed back to the time domain using the inverse FFT (iFFT) and added by a residual connection. After further refinement by a Conv, the output of the frequency-domain branch is concatenated with that of the time-domain branch at the channel dimension, and their concatenation is then fused using a Conv to obtain the final dual-domain feature. For an input feature $\bm{X}$, the output feature $\bm{Y}$ of DDA block is calculated as
\begin{equation}
\bm{X}_1=LReLU(Conv(\bm{X})),\bm{X}_2=Conv(LReLU(Conv(\bm{X}))),
\end{equation}
\begin{equation}
\bm{Y}_1=Conv(iFFT(LReLU(Conv(FFT(\bm{X}_1))))+\bm{X}_1),
\end{equation}
\begin{equation}
\bm{Y}_2=Conv(LReLU(Conv(\bm{X}_2)))+\bm{X}_2,
\end{equation}
\begin{equation}
\bm{Y}=Conv(Concatenate(\bm{Y}_1,\bm{Y}_2)),
\end{equation}

\subsection{TPNN}
For some 3D seismic gathers, the amplitudes at different sampling points may vary substantially. In such cases, the most adopted linear normalization approach in interpolation tasks, i.e., dividing each element by the maximum absolute value of the entire volume (denoted as Abs-Max), may scales the existing late arrivals or weak reflections to extremely small values. This phenomenon can be highly detrimental to the loss calculation and the subsequent backpropagation process during training, because signals with extremely large amplitudes can dominate the loss calculation, making it difficult for weak reflections to be effectively recovered. Therefore, we propose TPNN strategy to replace Abs-Max. In TPNN, the first stage aims to compress the range of data using the arcsinh function. Its forward and inverse transformations are defined as follows:
\begin{equation}
\hat{\bm{x}}=arcsinh(\bm{x})=ln(\bm{x}+\sqrt{\bm{x}^2+1}),
\end{equation}
\begin{equation}
\bm{x}=sinh(\hat{\bm{x}})=\frac{\bm{e}^{\hat{\bm{x}}}-\bm{e}^{-\hat{\bm{x}}}}{2},
\end{equation}
where $\bm{x}$ denotes the input seismic data, and $\hat{\bm{x}}$ denotes the compressed seismic data. The adoption of the arcsinh function has the following advantages: 1) It is monotonically increasing, which can nonlinearly preserve the relative amplitude relationships in seismic data; 2) It is continuous and therefore differentiable everywhere, facilitating gradient propagation during training; 3) It eliminates the need for sign operations in log compression function employed by \cite{jin2021efficient}, thereby simplifying the compression process and reducing computational operations. Meanwhile, our preliminary experiments also showed that the training process became unstable when only range compression was applied, motivating us to adopt Z-score standardization as a second-stage normalization step in TPNN. The Z-score standardization and its inverse transformation are performed by
\begin{equation}
\bm{x}^\prime=\frac{\hat{\bm{x}}-\mu}{\sigma},
\end{equation}
\begin{equation}
\hat{\bm{x}}=\sigma\bm{x}^\prime+\mu,
\end{equation}
where $\bm{x}^\prime$ is the final normalized seismic data through TPNN, and $\mu$ and $\sigma$ denote the mean and standard deviation of $\hat{\bm{x}}$, respectively.

When TPNN is adopted, an entire 3D seismic volume is first normalized through TPNN, and then patched into 3D patches of shape 32×32×8. Next, traces are randomly masked to generate complete-incomplete data pairs, so as to establish training and validation datasets. Subsequently, they are fed into the LMGM network for training. The LMGM network is finally optimized in the TPNN domain. As for the test data, it should first be TPNN-normalized before being input into the trained model, following which inverse TPNN normalization (iTPNN) is conducted to obtain the final processed data.

\section{Experiments}
In this section, we take interpolation as an example to validate the effectiveness and efficiency of the proposed LMGM framework.

\subsection{Environments and hyperparameters settings}
The training and testing programs are executed on Pytorch 2.10.0 and CUDA 12.4, running on an Ubuntu 22.04 operation system. Mamba v2.3.1 is installed from \url{https://github.com/state-spaces/mamba/releases#release-v2.3.1.} Hardware configurations consist of an NVIDIA A800 GPU with 80GB of memory, an Intel(R) Xeon(R) Gold 6348 CPU with 28 cores at 2.6GHz frequency, and 120GB RAM. Taking into account performance, computational resources, and fair comparison, we set the hyperparameters as shown in Table \ref{tab1}.
\begin{table}[htbp]
\centering
\caption{Hyperparameters of LMGM.}
\label{tab1}
\begin{tabular}{lc}
\toprule
\textbf{Hyperparameters} & \textbf{Specifications} \\
\midrule
Optimizer                   & AdamW\citep{loshchilov2017decoupled} \\
Loss function               & L2-norm \\
Data patch size             & 32×32×8 \\
Batch size                  & 4 \\
Number of training epochs   & 20 \\
Initial Learning rate       & 10\textsuperscript{-4} \\
Scheduler                   & Cosine Annealing \citep{loshchilov2016sgdr}\\
Input channel number        & 1 \\
Embedding channel number    & 96 \\
\bottomrule
\end{tabular}
\end{table}

\subsection{Metrics and competitive methods}
In this study, the signal-to-noise ratio (SNR) is used to quantitatively evaluate the performance of different methods. For a 3D ground truth $\bm{x}$, its SNR is defined as
\begin{equation}
SNR=10{log}_{10}(\frac{{||\bm{x}||}_2^2}{{||\bm{n}||}_2^2}),
\end{equation}
where ${||\cdot||}_2$ denotes the L\textsubscript{2}-norm, and $\bm{n}$ represents the noise. In this section, since the interpolation is used as an example, the difference between complete data and interpolated result $\bm{y}$, i.e., $(\bm{x}-\bm{y})$, is used in the denominator to replace the noise $\bm{n}$.

We select an CNN-based artificial neural network (ANN; \citealp{chai2020deep}) and a full-attention Transformer-based network (FAT) as competitive methods. The former is an encoder-decoder U-Net \citep{ronneberger2015u} structure extended to 3D seismic data. The latter is obtained by replacing the 3DM blocks with Pre-LN Transformer encoders \citep{xiong2020layer}.

\subsection{Synthetic example}
\subsubsection{Data preparation and training}
\begin{figure}[htbp]
    \centering
    \includegraphics[width=1\textwidth]{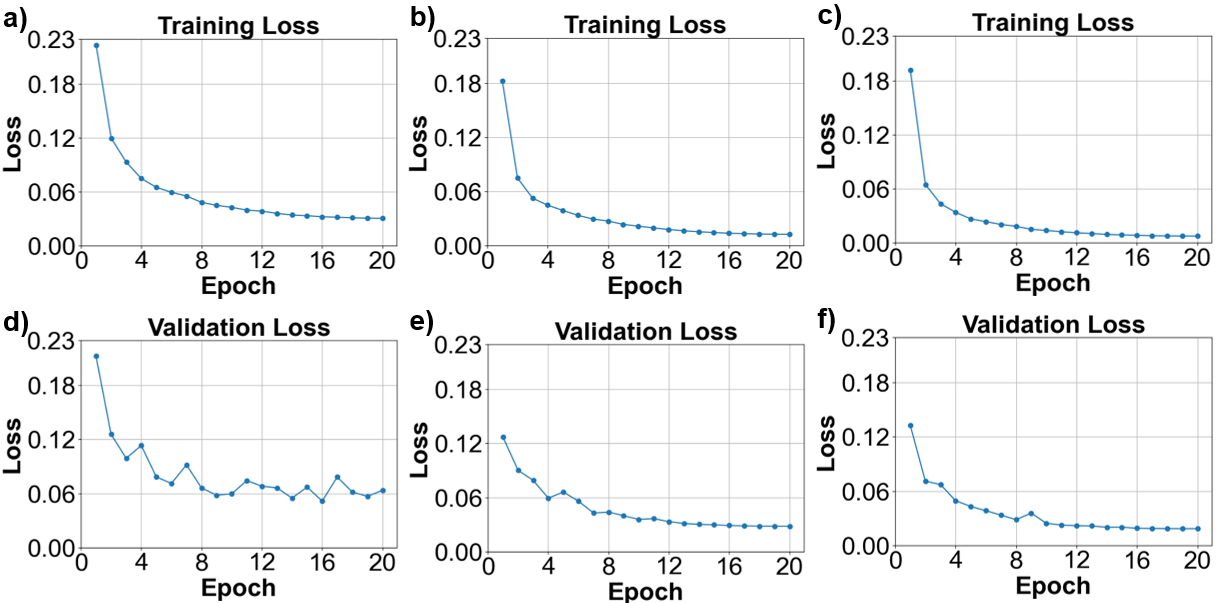}
    \caption{L\textsubscript{2} loss curves on SEG C3 45 shots dataset. (a–c) Training loss curves of ANN, FAT and LMGM, respectively; and (d–f) the corresponding validation loss curves.}
    \label{fig4}
\end{figure}

We first use SEG C3 45 shots dataset, an open dataset available at \url{ https://wiki.seg.org/wiki/SEG_C3_45_shot}, which contains 45 shot gathers, with 0.008 s sample rate and 625 sampling points per trace. Each shot gather corresponds to the same 201×201 receiver grid, with a spatial interval of 20 m in both the inline and crossline directions. We use the first three shot gathers to generate datasets. First, they are normalized through TPNN. Then, we generate sufficient 3D data patches for training and validation by leveraging a sliding window technique, which moves among the three dimensions at specific strides and allows overlapping. Next, for LMGM and FAT, we generate a training dataset, where the extracted 3D patches is of shape 32×32×8; for ANN, to satisfy the minimum data size requirement, we generate another training dataset with patches of shape 32×32×32. Both datasets contain 6000 3D patches for training and 1500 for validation. We randomly removed 40\%–60\% of the traces from each patch to generate complete-corrupted pairs. Subsequently, three models of the corresponding three methods are trained by feeding the data patch pairs into the three networks. Fig. \ref{fig4} displays the L\textsubscript{2} loss curves for both training and validation of the three methods.

\subsubsection{Testing results}
\begin{figure}[htbp]
    \centering
    \includegraphics[width=1\textwidth]{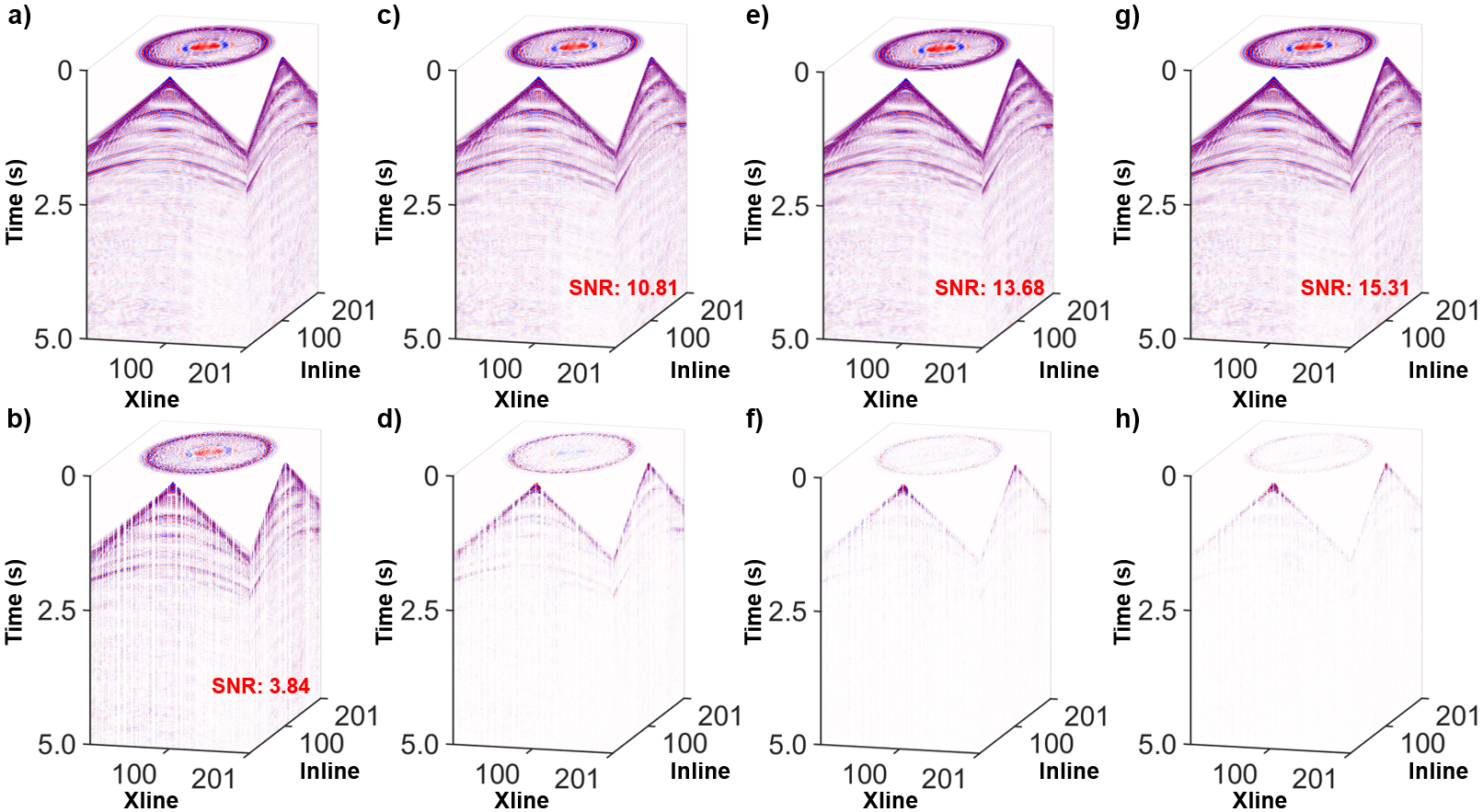}
    \caption{Interpolation on the 25th shot of the SEG C3 45 shots dataset. (a) Complete data, (b) 50\% randomly sampled data, (c, e, and g) interpolated results of ANN, FAT and LMGM, respectively; and (d, f, and h) the corresponding difference records.}
    \label{fig5}
\end{figure}

We select the entire 25th shot as the test data as shown in Fig. \ref{fig5}a, with a shape of 625×201×201. Corrupted data is generated by randomly removing 50\% of the total traces, as illustrated in Fig. \ref{fig5}b. We use TPNN to normalize the test data, and input it into the three trained models. To ensure a fair comparison among the three methods, we also adopt a sliding window technique during testing, which is similar to that used for training/validation dataset generation. The sliding window is moved with a stride equal to 50\% of the 3D patch size (16×16×4 for LMGM and FAT, and 16×16×16 for ANN), and predictions in overlapping regions were averaged to mitigate the edge effects. After the results are generated from the model, iTPNN operation is applied to obtain the final interpolated results, as shown in Fig. \ref{fig5}c, \ref{fig5}e and \ref{fig5}g along with their SNR values. Fig. \ref{fig5}d, \ref{fig5}f and \ref{fig5}h present the difference records between the interpolated results and the complete data. As shown in Fig. \ref{fig5}c, ANN demonstrate the poorest performance among the three methods because it can only model local correlation. In comparison, FAT and LMGM achieve a substantially higher level of interpolation performance, and LMGM recovers the most continuous events with highest numerical result in Fig. \ref{fig5}g and least signal leakage in Fig. \ref{fig5}h.

\begin{figure}[htbp]
    \centering
    \includegraphics[width=1\textwidth]{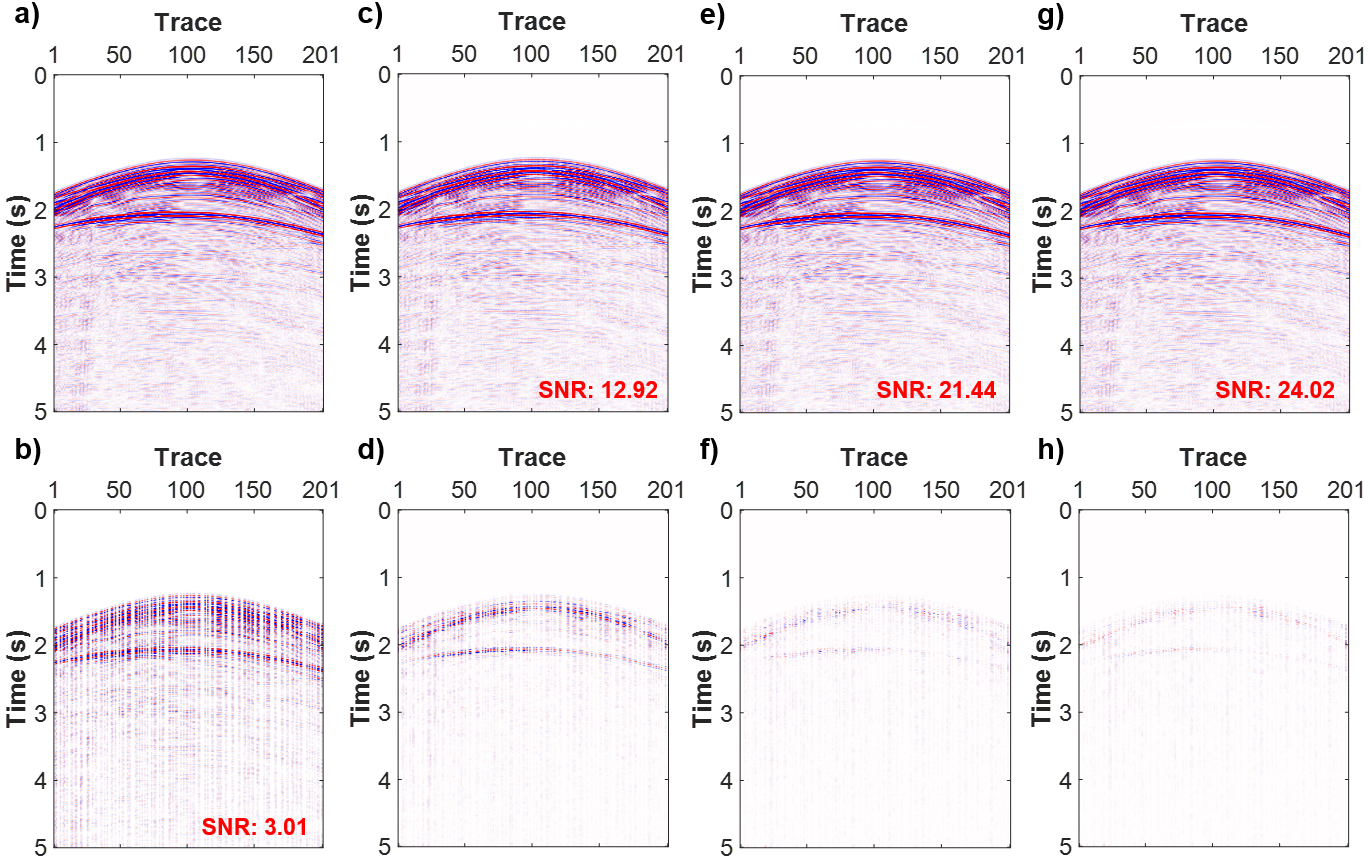}
    \caption{Comparisons of the second line extracted from Fig. \ref{fig5}. (a) Complete data, (b) Corrupted data, (c, e, and g) interpolated results of ANN, FAT and LMGM, respectively; and (d, f, and h) the corresponding difference records.}
    \label{fig6}
\end{figure}

The second line from each subplot in Fig. \ref{fig5} is extracted and presented in Fig. \ref{fig6} for a clearer comparison. As shown in Fig. \ref{fig6}c, the result obtained by ANN exhibit obvious discontinuous events, along with the most severe signal leakage in Fig. \ref{fig6}d. Meanwhile, FAT can effectively recover the missing traces as in Fig. \ref{fig6}e. In comparison, LMGM produces the best result in Fig. \ref{fig6}g.

\begin{figure}[htbp]
    \centering
    \includegraphics[width=1\textwidth]{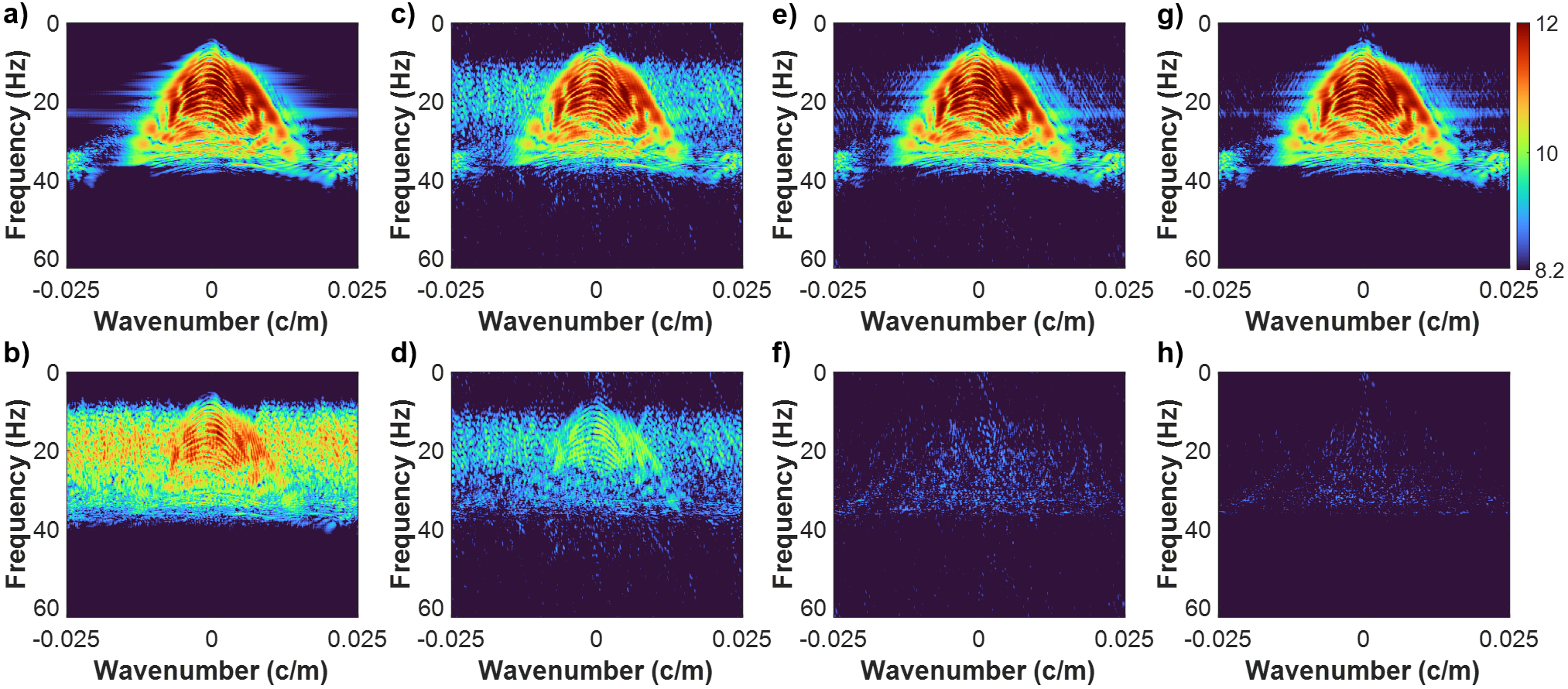}
    \caption{f-k spectra of the (a) complete data (Fig. \ref{fig6}a), (b) Corrupted data (Fig. \ref{fig6}b), (c, e, and g) interpolated results by ANN, FAT and LMGM (Fig. \ref{fig6}c, \ref{fig6}e, and \ref{fig6}g), respectively; and (d, f, and h) the corresponding difference records (Fig. \ref{fig6}d, \ref{fig6}f and \ref{fig6}h).}
    \label{fig7}
\end{figure}

Furthermore, in Fig. \ref{fig7}, we plot the f-k spectra of the complete data (Fig. \ref{fig6}a), Corrupted data (Fig. \ref{fig6}b), the three interpolated results (Fig. \ref{fig6}c, \ref{fig6}e, and \ref{fig6}g), and the corresponding difference records (Fig. \ref{fig6}d, \ref{fig6}f, and \ref{fig6}h). As shown in Fig. \ref{fig7}b, the missing traces introduce severe aliasing energy into the f-k spectrum, substantially degrading its spectral integrity. Strong residual interference exists in the result of ANN as illustrated in Fig. \ref{fig7}c. Although FAT alleviates this issue, its energy leakage in Fig. \ref{fig7}f remains more pronounced than that of LMGM in Fig. \ref{fig7}h. By contrast, the recovered spectrum of LMGM is the closest to the ground truth among the three methods. In summary, the above comparisons validate the effectiveness of LMGM on the synthetic dataset.

\subsection{Field example}
\subsubsection{Data preparation and training}
\begin{figure}[htbp]
    \centering
    \includegraphics[width=1\textwidth]{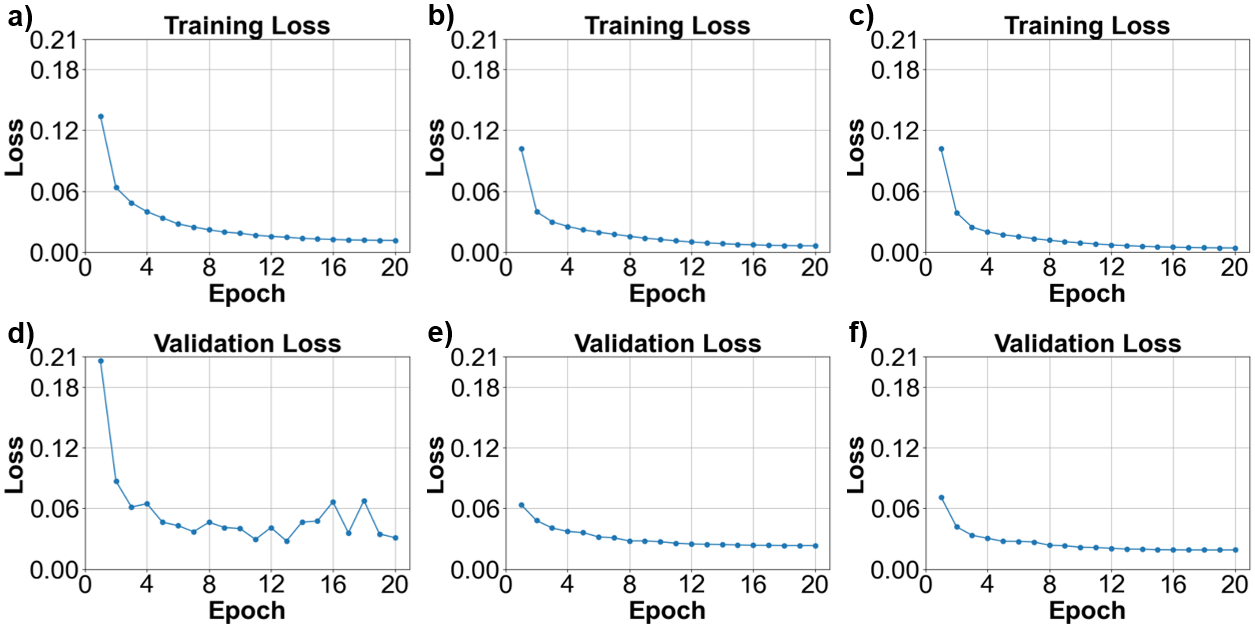}
    \caption{L\textsubscript{2} loss curves on field marine dataset. (a–c) Training loss curves of ANN, FAT and LMGM, respectively; and (d–f) the corresponding validation loss curves.}
    \label{fig8}
\end{figure}

For the evaluation on field data, we select marine seismic data from a survey area in Beibu Gulf, China. Specifically, we use 50 consecutive shot gathers, with the first 30 gathers used for training and the 40th gather used for testing. The time sampling rate is 0.001 s. Each shot gather contains 1000 sampling points, 128 crossline points and 4 inline points. Due to the limited number of lines, we pad them with zeros along the inline dimension before performing TPNN, so as to meet the requirements of all three methods. Training and validation datasets for ANN, FAT, and LMGM are generated in the same manner as for the synthetic data. Fig. \ref{fig8}a–\ref{fig8}f show the convergence behavior of the training and validation curves of the three methods on the field marine dataset.

\subsubsection{Testing results}
\begin{figure}[htbp]
    \centering
    \includegraphics[width=1\textwidth]{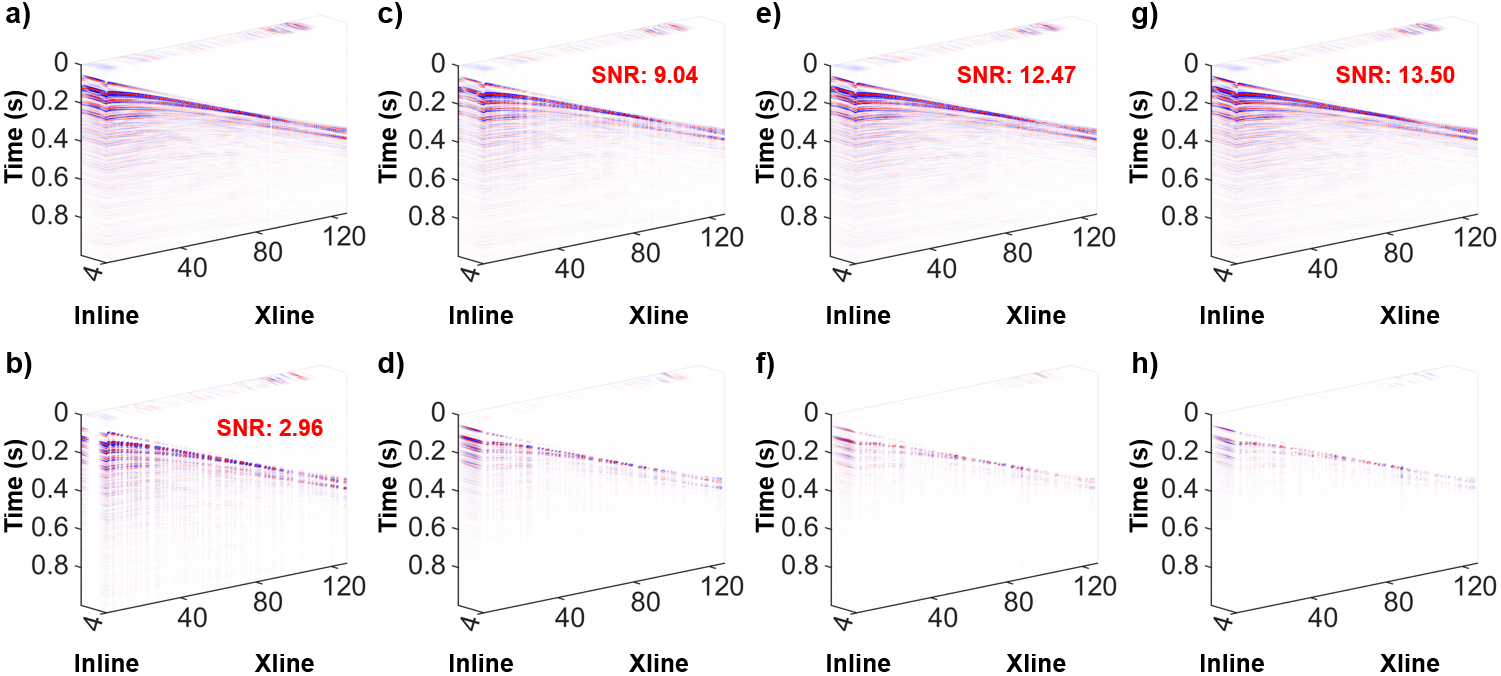}
    \caption{Interpolation on the 40th shot of the field marine dataset. (a) Complete data, (b) 50\% randomly sampled data, (c, e, and g) interpolated results of ANN, FAT and LMGM, respectively; and (d, f, and h) the corresponding difference records.}
    \label{fig9}
\end{figure}

The 3D comparison of the interpolation results is displayed in Fig. \ref{fig9}. As shown in Fig. \ref{fig9}c, \ref{fig9}e, and \ref{fig9}g, the performance of ANN and FAT remains markedly inferior to that of LMGM, which can be further supported by the more signal leakage in Fig. \ref{fig9}d and \ref{fig9}f compared to Fig. \ref{fig9}h.

\begin{figure}[htbp]
    \centering
    \includegraphics[width=1\textwidth]{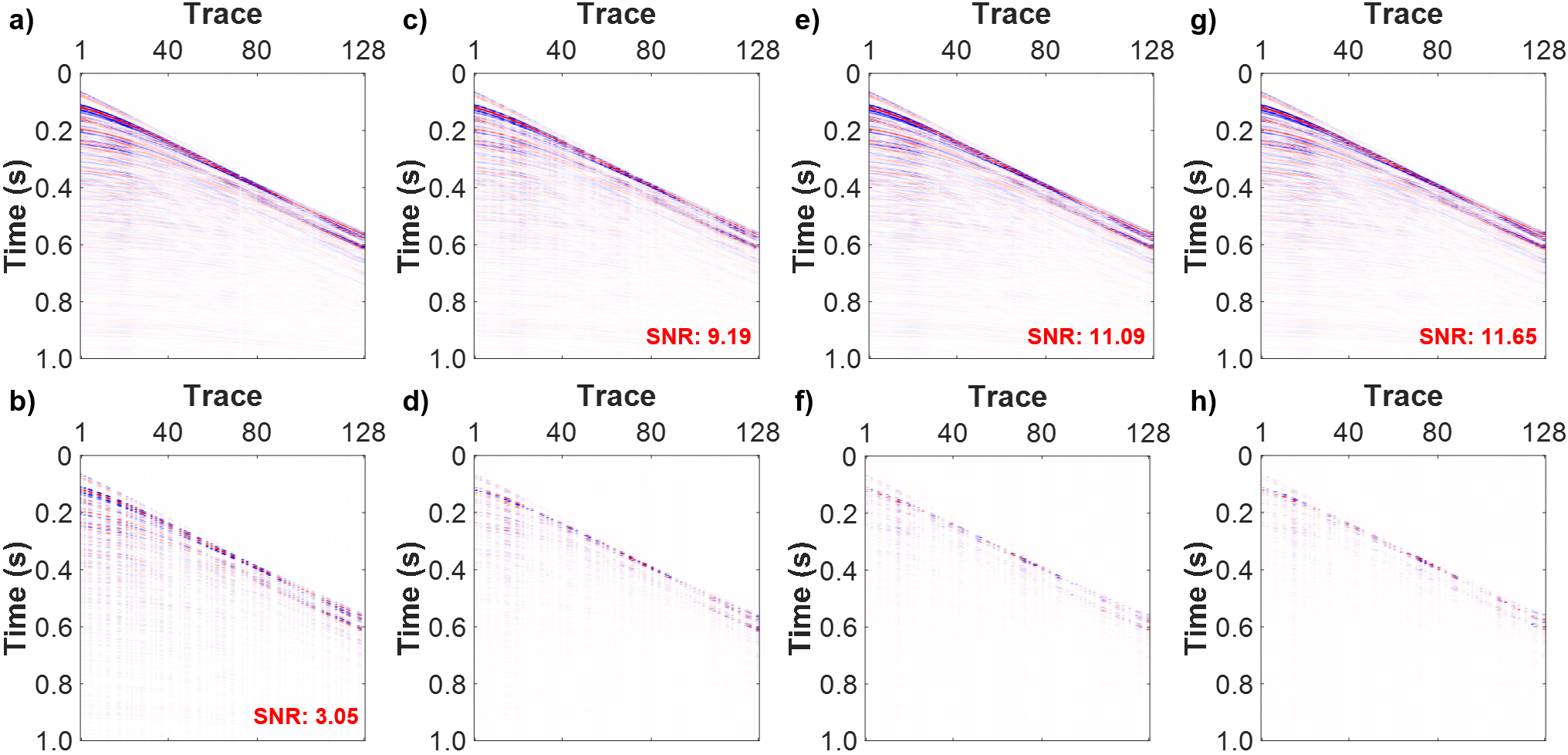}
    \caption{Comparisons of the first line extracted from Fig. \ref{fig9}. (a) Complete data, (b) Corrupted data, (c, e, and g) interpolated results of ANN, FAT and LMGM, respectively; and (d, f, and h) the corresponding difference records.}
    \label{fig10}
\end{figure}

Meanwhile, the first line is plotted in 2D, as shown in Fig. \ref{fig10}. It can be observed that the events recovered by LMGM are more continuous and smoother than that by ANN and FAT. Besides, the highest SNR value in Fig. \ref{fig10}g and the least leakage in Fig. \ref{fig10}h further indicate the great interpolation ability of LMGM.

\begin{figure}[htbp]
    \centering
    \includegraphics[width=1\textwidth]{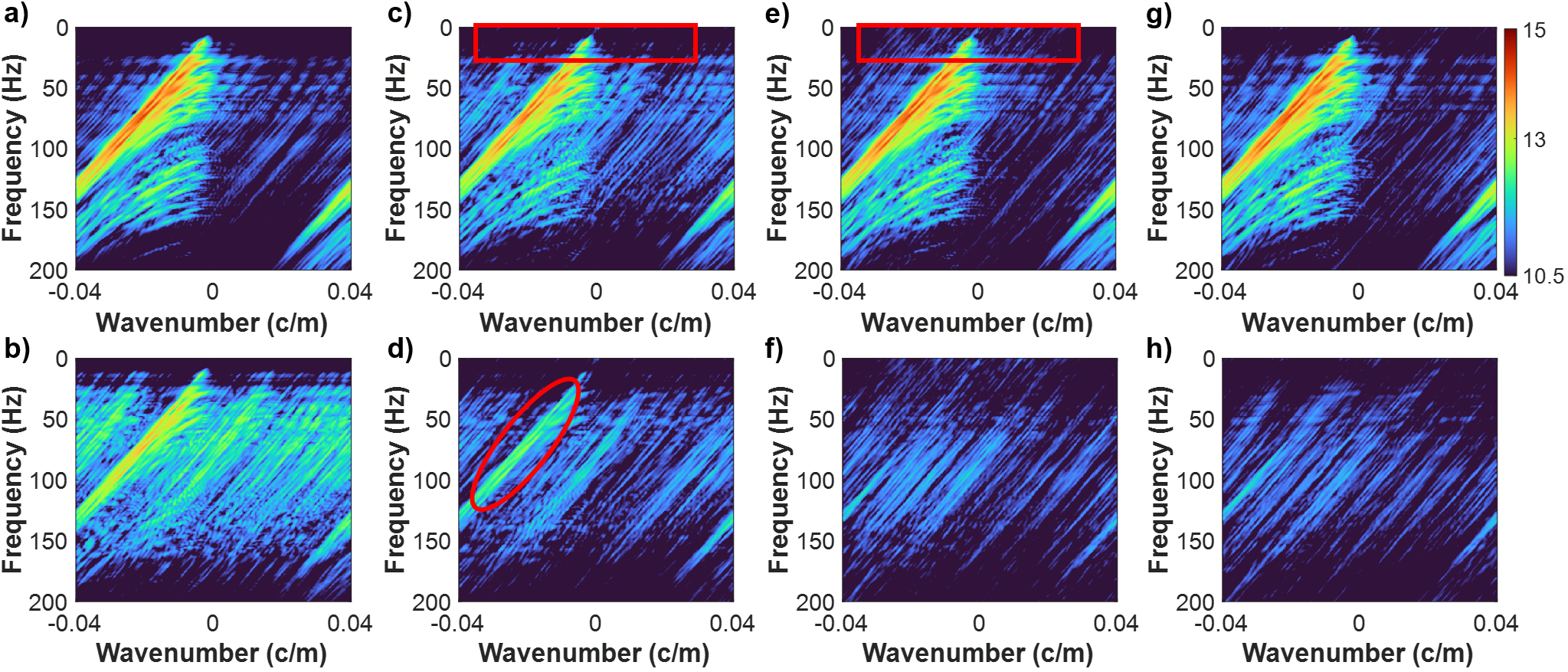}
    \caption{f-k spectra of the (a) complete data (Fig. \ref{fig10}a), (b) Corrupted data (Fig. \ref{fig10}b), (c, e, and g) interpolated results by ANN, FAT and LMGM (Fig. \ref{fig10}c, \ref{fig10}e, and \ref{fig10}g), respectively; and (d, f, and h) the corresponding difference records (Fig. \ref{fig10}d, \ref{fig10}f and \ref{fig10}h).}
    \label{fig11}
\end{figure}

Fig. \ref{fig11} also provides the comparison of Fig. \ref{fig10} in the f-k domain. As shown in Fig. \ref{fig11}c and \ref{fig11}e, the spectra recovered by ANN and FAT contain certain residual interference, as indicated by the red boxes. Meanwhile, Fig. \ref{fig11}d exhibits severe energy leakage as indicated by the red oval. In comparison, the f-k spectrum in Fig. \ref{fig11}g matches most closely with the ground truth in Fig. \ref{fig11}a, demonstrating the strong feature extraction ability of LMGM. In summary, LMGM exhibits great GM ability on the field marine dataset.

\subsection{Analysis of the computational cost}
One of the most important characteristics of LMGM is its lightweight design. To substantiate this claim, we quantitatively analyze the computational cost of the three different methods in this subsection. Since the experimental settings and dataset generation processes are basically the same for both the synthetic and field datasets, we have recorded the peak GPU memory usage and training time of the former as an example in Table \ref{tab2}. It is observed that ANN requires the least GPU memory and spend the least time on training, but these costs are slightly lower than that of LMGM. Meanwhile, ANN exhibits the poorest performance. By contrast, LMGM requires only approximately one-fifth of the memory and one-third of the training time of FAT. Overall, LMGM achieves the best processing performance at a substantially lower computational cost, striking a good balance between effectiveness and efficiency.
\begin{table}[htbp]
\centering
\caption{Computational costs of different methods on the synthetic dataset.}
\label{tab2}
\begin{tabular}{lcc}
\toprule
\textbf{Method} & \textbf{Peak GPU Memory Usage (MB)} & \textbf{Training Time (h)} \\
\midrule
ANN     & 3582  & 0.46\\
FAT     & 24372 & 2.60\\
LMGM    & 4764  & 0.97\\
\bottomrule
\end{tabular}
\end{table}

\section{Discussions}
\subsection{Ablation study}
In this subsection, we conduct ablation experiments to investigate the contributions of the DDA blocks and TPNN.

1) \textit{Contribution of the DDA blocks}: To analyze the effectiveness of the DDA blocks in the TPNN domain, we conduct two additional experiments: a) Keeping all other experimental settings unchanged, we directly remove the DDA blocks from the network, denoted as \textit{Abl1}; 2) Keeping all other experimental settings unchanged, we replace the 3DM block with two repeated Pre-LN structures, i.e., each DDA block is replaced with an additional “LN-3DSM-LN-MLP” structure, denoted as \textit{Abl2}. Both settings are trained from scratch, and their trained models are applied to test data in Fig. \ref{fig5}b. The comparisons of these models are provided in Table \ref{tab3}, including SNR of the interpolated result, peak GPU memory usage, and training time. Although \textit{Abl1} has slightly lower computational cost, but its SNR value has reduced by 0.52 dB, demonstrating the necessity of DDA blocks. As for \textit{Abl2}, with much higher computational cost, the SNR even decreased rather than improved, indicating that simply increasing the number of 3DSM blocks does not necessarily improve performance or replicate the role of the DDA blocks. In summary, DDA blocks have a positive effect on enhancing the global features in the frequency domain.
\begin{table}[htbp]
\centering
\caption{Contribution of the DDA blocks.}
\label{tab3}
\begin{tabular}{
l
    >{\centering\arraybackslash}p{0.2\linewidth}
    >{\centering\arraybackslash}p{0.3\linewidth}
    >{\centering\arraybackslash}p{0.2\linewidth}
}
\toprule
\textbf{Method} & \textbf{SNR (dB)}& \textbf{Peak GPU Memory Usage (MB)} & \textbf{Training Time (h)} \\
\midrule
LMGM          & 15.31  & 4764  & 0.97\\
\textit{Abl1} & 14.79  & 4304  & 0.78\\
\textit{Abl2} & 15.10  & 7658  & 1.32\\
\bottomrule
\end{tabular}
\end{table}

\begin{figure}[htbp]
    \centering
    \includegraphics[width=0.8\textwidth]{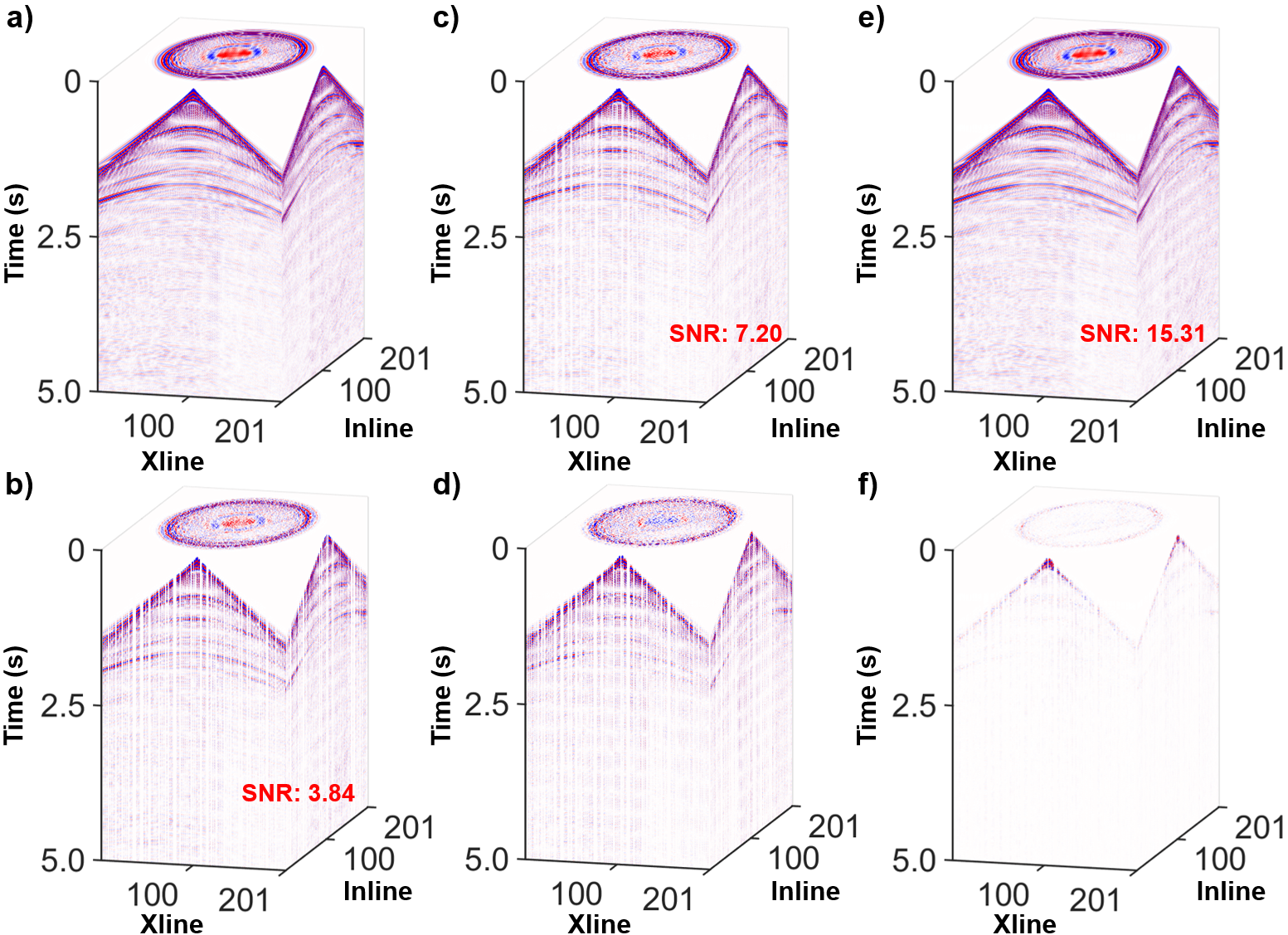}
    \caption{Comparisons between the Abs-Max normalization and TPNN. (a) Complete data, (b) 50\% randomly sampled data, (c and e) interpolated results of LMGM using Abs-Max normalization and TPNN, respectively; and (d and f) the corresponding difference records.}
    \label{fig12}
\end{figure}

2) \textit{Contribution of the TPNN}: As introduced above, TPNN is a strategy specifically designed for recovering data with large amplitude variations. We observed that the maximum absolute amplitude of the synthetic data in Fig. \ref{fig5}a reaches 1.043×10\textsuperscript{4}, spanning four orders of magnitude. This may cause severe problems using Abs-Max normalization. To validate this point and demonstrate the necessity of TPNN for such data, we kept all other experimental settings unchanged and retrained the model using Abs-Max normalization for the synthetic data. Notably, the trained model also requires the test data to be Abs-Max-normalized. The interpolation results obtained using Abs-Max and TPNN normalizations are compared in Fig. \ref{fig12}, where the model trained by Abs-Max normalized dataset essentially fails to perform effective interpolation. To summarize, these comparisons validate the necessity of TPNN for recovering such data.

\subsection{Interpolation on post-stack seismic data}
\begin{figure}[htbp]
    \centering
    \includegraphics[width=1\textwidth]{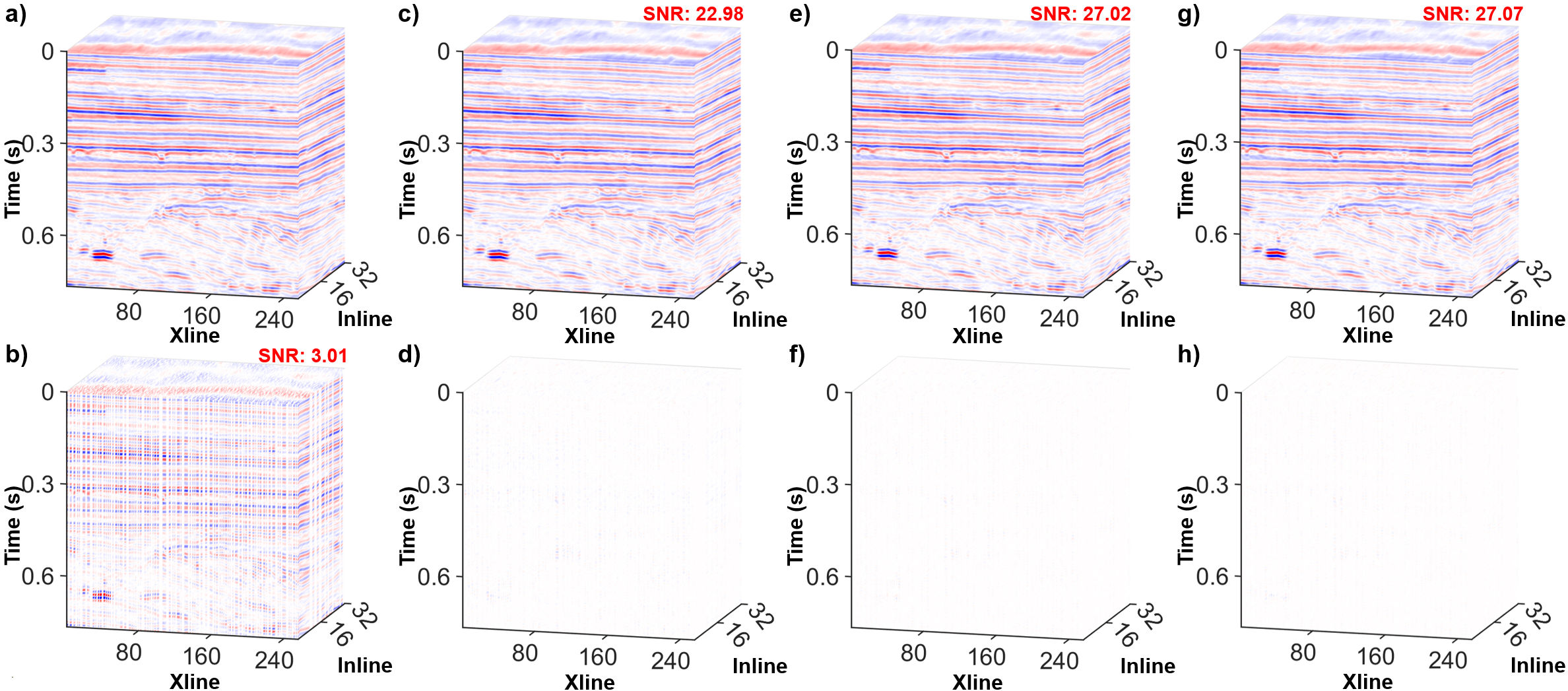}
    \caption{Interpolation on the Parihaka dataset. (a) Complete data, (b) 50\% randomly sampled data, (c, e, and g) interpolated results of ANN, FAT and LMGM, respectively; and (d, f, and h) the corresponding residual images.}
    \label{fig13}
\end{figure}

In the pre-stack data used in our previous experiments, substantial amplitude variations may occur. In contrast, the energy distribution of post-stack data is generally more uniform. To evaluate the performance of LMGM on post-stack data, we selected the Parihaka dataset (available at \url{https://wiki.seg.org/wiki/Parihaka-3D}) for training and testing. We employ Abs-Max normalization instead of TPNN for this dataset.  The time sampling interval is 0.003 s. We extract a 3D volume with a shape of 256×650×186, with the first 154 lines for training dataset generation and the last 32 lines for testing. The results and their corresponding SNR values are presented in Fig. \ref{fig13}. Through observation, LMGM continues to achieve the best interpolation performance on the post-stack data with the highest SNR value and the least signal leakage, which is consistent with the trend observed in the pre-stack experiments. Overall, these results further demonstrate the effectiveness of LMGM on post-stack data.

\subsection{Application of LMGM to Seismic Denoising}
\begin{figure}[htbp]
    \centering
    \includegraphics[width=1\textwidth]{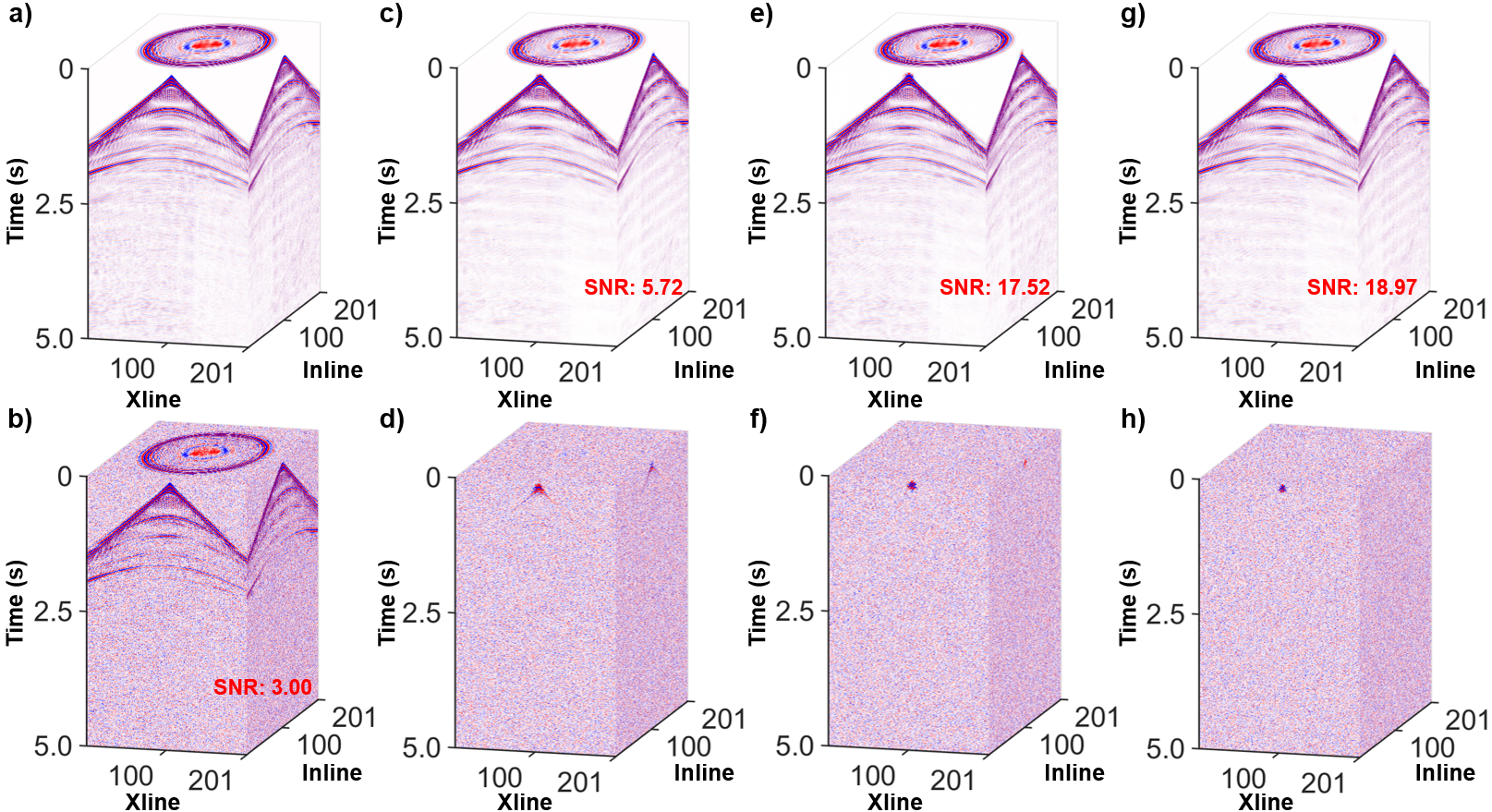}
    \caption{Denoising on the 25th shot of the SEG C3 45 shots dataset. (a) Clean record, (b) noisy record, (c, e, and g) denoised results of ANN, FAT and LMGM, respectively; and (d, f, and h) the corresponding predicted noise.}
    \label{fig14}
\end{figure}

To demonstrate the potential of LMGM as a general GM framework, we selected another common task, random noise attenuation, to further evaluate its performance. Specifically, keeping other experimental settings unchanged, we replaced the L\textsubscript{2}-norm loss function with the L\textsubscript{1}-norm, which is more commonly used for denoising tasks. Meanwhile, we generate data patches using the first three shots of the SEG C3 45 shot dataset without any normalization. This is because we noticed that the energy of random noise is relatively low and uniformly distributed. Applying either Abs-Max normalization or TPNN will make the noise amplitude excessively small, thereby making it difficult for the network to accurately separate the noise from signals. We added Gaussian noise to the first three shot gathers to reduce their SNR to 3.00 dB, followed by sliding window-based 3D patching to generate clean-noisy pairs.  Patches of size 32×32×8 were used for FAT and LMGM, whereas patches of size 32×32×32 were used for ANN. Both datasets contain 6000 3D patch pairs for training and 1500 for validation. Through training, three models of the corresponding three methods are obtained. Their denoising performance are evaluated on the 25th shot in Fig. \ref{fig14}. In Fig. \ref{fig14}b, we added Gaussian noise to reduce its SNR to 3.00 dB, where the presence of strong random noise visually overwhelms and masks the deep reflection signals. ANN can effectively attenuate random noise, but the preservation of the first-arrival signals is the poorest among the three methods, as shown in the predicted noise in Fig. \ref{fig14}d. FAT improves this issue in Fig. \ref{fig14}f; however, it still exhibits more signal leakage than LMGM in Fig. \ref{fig14}h. In summary, LMGM achieves remarkable denoising performance with the highest numerical result and least signal leakage among the three methods. Nevertheless, we can still observe artifacts around the first-arrival signals in Fig \ref{fig14}c, \ref{fig14}e, and \ref{fig14}g. This issue may be related to the dataset generation strategy, as too few patches cover this region, preventing the networks from sufficiently learning this type of signal.

\section{Conclusion}
In this paper, we propose a general lightweight Mamba-based GM (LMGM) framework for 3D seismic data processing. LMGM captures long-range dependencies among the three dimensions of 3D seismic data through 3DSM blocks. To further enhance feature representations of the extracted global features, we design the DDA block to enable more effective utilization of the frequency-domain correlation. For recovering some data with large amplitude variations, we propose TPNN to balance the amplitude of seismic data. Experimental results demonstrate that the proposed LMGM framework outperforms the two competitive methods, ANN and FAT, in both interpolation and denoising tasks. Comparisons of computational costs indicate the high efficiency of LMGM. In conclusion, LMGM is a lightweight yet effective approach for 3D seismic data processing.

Our proposed framework still has several limitations. First, the DDA block does not perform GM in the frequency domain. Second, TPNN is designed as a plug-and-play strategy, making it useful but not necessarily required for all tasks. Finally, a dedicated model was retrained for each dataset and task, indicating that the generalization of our method across different data and tasks still requires improvement. In future work, we will investigate highly generalizable foundation models for seismic data to overcome these limitations.

%

\section{Conflict of Interest}
The authors declare no conflict of interest.

\section{Data and Materials Availability}
The source code for dataset preparation, network training and testing is available by contacting the authors. The SEG C3 45 shots dataset and the Parihaka dataset associated with this research are available at \url{https://wiki.seg.org}. The field marine dataset associated with this research is confidential and cannot be released.
%

\bibliographystyle{plainnat} 
\bibliography{references}

\end{document}